\documentclass[twocolumn,amsmath,amssymb,nofootinbib]{revtex4}
\usepackage{graphicx}
\usepackage{dcolumn}
\usepackage{color}
\usepackage{bm}
\usepackage{amsmath,amssymb,graphicx}
\usepackage{epsfig}
\usepackage{enumerate}
\usepackage{array}
\usepackage{multirow}

\begin{document}
\title
{Quantized Area of the Schwarzschild Black Hole: \\
A non-Hermitian Perspective}
\author{Bijan Bagchi\footnote{E-mail: bbagchi123@gmail.com}$^1$,  Aritra Ghosh\footnote{E-mail: ag34@iitbbs.ac.in}$^2$, Sauvik Sen\footnote{E-mail: sauviksen.physics@gmail.com}$^3$}

\vspace{2mm}

\affiliation{$^{1}$Brainware University,
Barasat, Kolkata, West Bengal 700125, India\\
$^{2}$School of Basic Sciences, Indian Institute of Technology Bhubaneswar, Jatni, Khurda, Odisha 752050, India\\
$^3$ Department of Physics, Shiv Nadar Institution of Eminence, Gautam
Buddha Nagar, Uttar Pradesh 203207, India}

\vskip-2.8cm
\date{\today}
\vskip-0.9cm


\begin{abstract}
In this work, our aim is to link Bekenstein’s quantized form of the area of the event horizon to the Hamiltonian of the non-Hermitian Swanson oscillator which is known to be $\mathbb{PT}$-symmetric. We achieve this by employing a similarity transformation that maps the non-Hermitian quantum system to a scaled harmonic oscillator. Our procedure is standard and well known. We, first of all, consider the unconstrained reduced Hamiltonian which is directly expressed in terms of the Schwarzschild mass and implies a periodic character for the conjugate momentum (which represents the asymptotic time coordinate), the period being the inverse Hawking temperature. This leads to the quantization of the event-horizon area in terms of the harmonic-oscillator levels. Within the framework of the Swanson oscillator, we proceed to derive novel expressions for the Hawking temperature and the black hole entropy. Notably, the logarithmic area-correction term $-(1/2)\ln$(area) is consistent with our results whereas $-(3/2)\ln$(area) is not. 
\end{abstract}

\maketitle

\section{Introduction}

As is well known, the idea of `event-horizon quantization' of a black hole was initially uncovered by Bekenstein \cite{bek1, bek2} who had conjectured that the area $A$ of the event horizon could be quantized in the manner $ \Delta A = \kappa l_P^2$, where the constant of proportionality $\kappa$ is some number of order unity. Subsequently, Bekenstein and Mukhanov gave a precise determination of it in the form \cite{bek4,bek3}
\begin{equation}\label{4}
    A = 16 \pi \left (\frac{GM}{c^2} \right )^2,
\end{equation} and argued on the basis of adiabatic invariants, for a uniform spacing of $A$ similar to the Bohr-Sommerfeld
quantization rule, and in consequence for discrete values for $M$. This implies that as the black hole evaporates \cite{lou}, transitions are only possible between mass states of discrete values whereby radiation occurs in chunks with an accompanying wavelength of similar order as the Schwarzschild radius, \(r_h = 2GM/c^2\). In the above, \(G\) is the Newton's constant and \(c\) is the speed of light in vacuum. In terms of the Schwarzschild radius, the Schwarzschild metric (in static coordinates) reads as
\begin{equation}\label{1.2}
    ds^2 = -\left (1-\frac{r_h}{r} \right ) c^2 dt^2 + \left (1-\frac{r_h}{r} \right )^{-1} dr^2 + r^2 d\Omega_2^2,
\end{equation}
where \(d\Omega_2^2 = d\theta^2 + \sin^2 \theta d\phi^2\) is the metric on a unit two-dimensional sphere. Notice that $M$ appears as a conserved charge, corresponding to the \(\partial_t\) Killing vector. \\

The idea of black holes radiating thermally at some non-zero temperature $T_{\rm H}$ was advanced by Hawking and his coworkers in the seventies \cite{haw1, haw2} indicating that black holes could decrease in mass (see, for example, \cite{silk}). The entropy of a black hole emitting a thermal stream of particles behaves like the area of its event horizon, as in
\begin{equation}\label{1.4}
    S_{\rm BH} = \frac{k_B c^3}{4 \hbar G} A,
\end{equation}
where $k_B$ is the Boltzmann constant and \(\hbar\) is the (reduced) Planck's constant. Of the other thermodynamic quantities that one can associate with a Schwarzschild black hole, the Hawking temperature turns out to be 
\begin{equation}\label{1.5}
    T_{\rm H} = \frac{\hbar c^3}{8 \pi k_B G M},
\end{equation}
showing that the temperature of the black hole is quantum in character, being proportional to $\hbar$. We remark that in the derivation of (\ref{1.5}), gravity was accounted for classically while non-gravitational fields were considered from a quantum point of view; this is familiar as semi-classical gravity. Such an approximation, however, breaks down when the dimension of the black hole reduces to Planck scale and the rules of quantum gravity take over (for example, see \cite{kiefer}). Quantum black holes exhibit some features like those of excited atoms \cite{kot, david, silk}. The clean-qubit simulation model has been used to demonstrate the similarity between the growth of quantum complexity and the volume of black holes \cite{pani}. 
Recent studies show that a discrete spectrum can have
observable imprints on the gravitational-wave signal emanating from an inspiraling, binary black hole \cite{datta}. However, a contrary view is also held in terms of so-called gravitational-wave echoes during black hole mergers \cite{coates}. \\

It may be remarked that general relativity takes the energy content of the universe to be smooth, down to the infinitesimal scales although in schemes of quantum gravity, by implementing certain altered forms of the uncertainty principle, a minimal-length criterion could be imposed to account for high-energy modifications \cite{bagchi1, bosso}. On the other hand, quantum mechanics, which accounts for the atomic and sub-atomic structure of matter, excludes the effects of gravity. A combination of these two apparently-disjoint fields of physics, if at all feasible \cite{bose, moller}, requires reconcilation at the extreme level of the Planck length $l_P$ defined, as usual, by the relation $l_P = \sqrt{\hbar G/c^3}$. \\

While analyzing the area spectrum of the Schwarzschild black hole, Louko and M\"{a}kel\"{a} \cite{lou} showed, by applying the tools of Hamiltonian quantization that the energy operator supports  discrete positive eigenvalues which are bounded from below (see also, \cite{mukh, deeg}). It means that transitions occur in lumps between quantized mass states with the accompanying radiation taking place in wavelengths of roughly the same order as the Schwarzschild radius. Obviously, it reflects a quantum-mechanical situation like that of an excited atom \cite{kot, david, silk}. The Hamiltonian form of the Einstein-Hilbert action, after employing an appropriate canonical transformation on the conjugate variables, namely, the mass $M$ and its conjugate momentum $P_M$ which obey the Poisson bracket $\{M, P_M\} = 1$, and subsequently eliminating the constraint, finally could be transformed to the form $\int (P_M \dot{M} - M)dt$. In this integral, the Hamiltonian is identified with $M$, and thus one can establish the equivalence of Schwarzschild mass with the Arnowitt-Deser-Misner (ADM) energy. The equations of motion imply $M$ to be a constant while $P_M$ plays the role of time: the latter is assumed to be a periodic variable that recurs after each interval of the inverse Hawking temperature $T_{\rm H}$. Making a formal correspondence of the canonical pair $(M, P_M)$ with the respective ones of the coordinate and momentum, and exploiting the periodic condition on the coordinate and momentum, the black hole area depicts the levels of a quantized oscillator \cite{mukh, jal}. We shall briefly review this in Sec. (\ref{revsec}) (see \cite{lou} for more details). \\

The aim of the present study is to analyze the area quantization of the event horizon using the machinery of non-Hermitian quantum mechanics, wherein, we will invoke the non-Hermitian Swanson oscillator to describe the quantized `area' levels. This allows for a re-computation of the thermodynamic quantities, namely, mass, temperature, and entropy of the system which now acquire novel modifications due to the area being described by a non-Hermitian system. We will work with the Hermitian representation of the Swanson oscillator; the latter is, in fact, pseudo-Hermitian\footnote{This ensures that the unitarity of the time evolution is maintained.}, allowing for such a representation to exist. We will compute the logarithmic corrections to the entropy and propose an upper bound for the magnitude of the coefficient characterizing the corrections. In the following, we shall adopt natural units $\hbar = c = k_B = 1$, in which \(l_P^2 = m_P^{-2} = G\). 

\section{Area quantization: A brief overview}\label{revsec}

The reduced action of a Schwarzschild black hole has the standard representation (see, for instance, \cite{kuc,das2,jal} and Appendix-(\ref{app})) that goes as
\begin{equation}\label{5}
    \mathcal{I} = \int \left (P_M\dot{M} - H (M) \right ) dt,
\end{equation}
where the pair $(P_M, M)$ represents a set of conjugate variables, the Hamiltonian $H$ is independent of $P_M$ and has the reduced form $H(M) = M$, $M$ being an integration constant. \\

Following the equations of motion generated by (\ref{5}), one finds $P_M = t$ up to a sign, i.e., $P_M$ plays the role of time. Thus, it is signified that
\begin{equation}\label{6}
    P_M \sim P_M + \frac{1}{T_{\rm H}}, \quad T_{\rm H} = \frac{m_P^2}{8\pi M},
\end{equation} where the Hawking temperature \(T_{\rm H}\) is introduced from the periodicity of the time variable. The above boundary condition verifies that there is no conical singularity in the two-dimensional Euclidean sector and that the `effective' phase space is a wedge that is cut out from the full $(M, P_M)$ phase space, bounded by the mass axis and the line $P_M=1/T_\text{H}$ \cite{Med}. Thus, from \cite{lou, das2}, one could 
make the following canonical transformation: $(M, P_M ) \rightarrow (x, p_x)$, which simultaneously opens up the phase space and also incorporates the periodicity condition (\ref{6}) as
\begin{align}
    x &= \sqrt{\frac{A}{4\pi G}} \cos(2\pi P_M T_{\rm H}), \label{7} \\
    p_x &= \sqrt{\frac{A}{4\pi G}} \sin(2\pi P_M T_{\rm H}),\label{8}
\end{align}
in which $A=4\pi r_h^2 = 16\pi G^2 M^2$ is the area of the event horizon. As such one can write\footnote{We have corrected a few typos in \cite{jal}.}
\begin{equation}\label{9}
    A = \frac{4\pi}{m_P^2} \left (p_x^2 + x^2 \right ),
\end{equation}
showing a striking resemblance with the harmonic-oscillator Hamiltonian $H = \frac{p_x^2}{2m} + \frac{1}{2} m \omega^2 x^2$, with mass $m$ and angular frequency $\omega$. In fact, with $m = \frac{1}{\omega} = \frac{m_P^2}{8\pi}$, the area $A$ exactly coincides with the Hamiltonian whose wave functions are the harmonic-oscillator states in the position representation.\\

Now, for this problem of a quantum oscillator, let us follow the treatment of \cite{jal} to outline below a few salient steps relevant for the description of area quantization. Here, the Hamiltonian operates upon a ket $|n \rangle$ to generate the eigenvalues $(n + \frac{1}{2})\omega$, where $n = 0, 1, 2,\cdots$. It then straightforwardly follows that the area of the event horizon supports a discrete spectrum of states (as in atoms) as evidenced by the spacings
\begin{equation}\label{10}
    \mathcal{A}\equiv A(n) = \left (n + \frac{1}{2} \right )\omega = \frac{8 \pi}{ m_P^2} \left (n + \frac{1}{2} \right ),
\end{equation} where \(n= 0, 1, 2, \cdots\). Since \(A \sim r_h^2\) and \(M \sim r_h\), it is in turn implied that the black hole mass is quantized \cite{mukh} and restricted to the discrete values
\begin{equation}\label{11}
   \mathcal{M} \equiv M (n) \approx m_P\sqrt{\frac{1}{2}\left (n+\frac{1}{2}\right )}, \quad n= 0, 1, 2, \cdots.
\end{equation}
We thus see that the accompanying emitted radiation would have an approximate frequency while a transition is made from level $n+1$ to level $n$, according to  
\begin{equation}\label{12}
    \omega_0 =  M(n+1)- M(n) \approx \frac{m_P}{2\sqrt{2n}} \approx \frac{m_P^2}{4 \mathcal{M}} \left (1+\frac{m_P^2}{8 \mathcal{M}^2} \right ),
\end{equation}
for large $n$. Note that the relevant timescale can be approximated by 
\begin{equation}\label{13}
    \tau_n^{-1} = \frac{|\dot{\mathcal{M}}|}{\omega_0} \approx \frac{4 \mathcal{M}|\dot{\mathcal{M}}|}{m_P^2}\left (1-\frac{m_P^2}{8 \mathcal{M}^2} \right ),
\end{equation}
where an overhead dot stands for the derivative with respect to the time. Actually,
from Stefan-Boltzmann law, we would have $|\dot{\mathcal{M}}| = \sigma_s A \mathcal{T}^4$. 
 With the width of each level $n$ being $|W_n| = \kappa \omega_0$, where the levels possess a well-defined structure for $\kappa <<1$, and the time $\tau_n$ obeys the relation $W_n \tau_n \approx 1$ \cite{mukh}, the temperature using (\ref{12}) and (\ref{13}) assumes the form
\begin{equation}\label{14}
    \mathcal{T}=\left(\frac{\kappa}{16\pi \sigma_S}\right)^{\frac{1}{4}} \frac{m_P^2}{2 \mathcal{M}}\left(1+\frac{m_P^2}{16 \mathcal{M}^2}\right),
\end{equation}
where $\sigma_S = \frac{\pi^2}{60}$ is the numerical value of the Stefan-Boltzmann constant. (\ref{14}) shows an $\mathcal{O}(\frac{m_P^4}{\mathcal{M}^3})$ modification to Hawking's estimate of the surface temperature of the black hole. It may be mentioned that calculations of Hawking temperature for a quantum-corrected black hole geometry already exist in the literature \cite{gango}.\\

The representation (\ref{14}) allows us to calculate the corresponding entropy $\mathcal{S}$ from the integral $\mathcal{S} = \int \frac{d\mathcal{M}}{\mathcal{T}}$. Taking $\kappa = \frac{\sigma_S}{16\pi^3}$, it readily turns out to be  

\begin{equation}\label{15}
    \mathcal{S} = S_{\rm BH} -\frac{\pi}{4}\ln S_{\rm BH} + \mbox{constant}, 
\end{equation}
where $S_{\rm BH} (= \frac{4\pi \mathcal{M}^2}{m_P^2})$ stands for the Bekenstein-Hawking form of the entropy. We should emphasize that the above result depends on a number of factors, the main one being the approximation we made to arrive at the first-order expressions of $\mathcal{M}, \omega_0, |W_n|$, and $\mathcal{T}$. The comforting point is that the logarithmic correction to the entropy as given in (\ref{15}) emerges in the right order of magnitude.\\

It may be pointed out that the relation (\ref{15}) can be translated into an area law with a logarithmic correction that goes as
\begin{equation}\label{16}
\mathcal{S} = \frac{m_P^2}{4}\mathcal{A} -\frac{\pi}{4}\ln (\mathcal{A}) + \mbox{constant}.
\end{equation}
One can compare it with the result of \cite{mitra}, where the logarithmic correction showed a coefficient of $\frac{1}{2}$. This was compared with other approaches that had coefficients either to be $\frac{1}{2}$ or $\frac{3}{2}$ which originated from quantum fluctuations of spacetime geometry when matter fields were absent in \cite{kaul}, but concluded in favor of the former coefficient. With this background, let us discuss the problem of area quantization from the point of view of the non-Hermitian Swanson oscillator. 

\section{Non-Hermitian Swanson oscillator: Known aspects}
There are various ways to incorporate `non-Hermiticity' in quantum mechanics (see, for instance, \cite{bend1, mos, zno1}).
 An elegant attempt was advanced by Swanson \cite{swa} who proposed a general form for a manifestly non-Hermitian oscillator expressed in terms of 
the traditional bosonic creation and annihilation operators, namely, $a^\dagger$ and $a$, respectively, obeying $[a, a^\dagger] = 1$. It reads
\begin{equation}\label{17}
 H^{(\alpha, \beta)} =  \left (H^{(\beta, \alpha)} \right)^\dagger = \omega a^\dagger a + \alpha a^2 + \beta {a^\dagger}^2    + \frac{1}{2} \omega,
\end{equation}
where $\omega, \alpha,$ and $\beta$ are three non-zero real coefficients (we will further take \(\omega > 0\)) such that $\alpha \neq \beta$ and $\Omega^2 = \omega^2 - 4\alpha \beta > 0$. Note that $a$ and $a^\dagger$ are connected to $x$ and $p$ by the relations $(a, a^\dagger) = \sqrt{\frac{m\omega}{2}} (x \pm \frac{ip}{m\omega})$ in which $m$ is the mass of the oscillator and $[x, p] = i$. An extended-oscillator system like (\ref{17}) can be completely solved using the standard operator techniques by employing a generalized Bogoliubov transformation \cite{swa}. The above Hamiltonian is $\mathbb{PT}$-symmetric for all
values of $\alpha$ and $\beta$ (but
Hermitian only if $\alpha = \beta$). This can be readily established by applying the transformations $\mathbb{P} : a \rightarrow -a$ and $\mathbb{T} : a \rightarrow a$. In this connection, we might mention that $\mathbb{PT}$-symmetric schemes basically address phase transitions between regions where it remains unbroken and the ones where it is not. As a consequence, the eigenvalues transit between real values to becoming complex \cite{bend1}. Apart from analytical evaluations, numerical algorithms for the diagonalization of $\mathbb{PT}$-symmetric Hamiltonians have also been developed \cite{noble}. In recent times, there has been an incredible amount of work in understanding the rich structure of $\mathbb{PT}$-symmetry in quantum systems (see \cite{bend2} for a thorough assessment). The guiding eigenstates of $ H^{(\alpha, \beta)}$ possess a real and positive set of discrete eigenvalues, a feature in accordance with the conjecture of Bender and Boettcher \cite{bend1}. A number of works are available in the literature on the Swanson Hamiltonian \cite{jones, bagchi2, mus, que, roy, bagchi3, fring, ortiz, rami}.  \\

In order to make connection with the black hole area, we first note that Jones \cite{jones} pointed out that a similarity transformation allows one to write down a Hermitian equivalent of $H^{(\alpha, \beta)} = H $. In the coordinate representation, and retaining the standard representation of $a$ and $a^\dagger$, it turns out to be that of a scaled harmonic oscillator, i.e., 
\begin{align}\label{18}
\begin{split}
h &= \rho H \rho^{-1} \\
  &= \frac{1}{2}(\omega-\alpha -\beta ) p_x^2 +\frac{1}{2}\frac{\Omega^2}{\omega -\alpha -\beta} x^2  ,
\end{split}
\end{align}
where $\rho=e^{\frac{1}{2}\lambda x^{2}}$, $\Omega^2=\omega^2-4\alpha\beta$, and we have replaced $x \rightarrow \frac{1}{\sqrt{m\omega}}x$. The accompanying eigenfunctions of $h$ in terms of normalization factors $N_n$ are
\begin{equation}\label{19}
\psi_{n}(x)= N_{n} e^{-\frac{1}{2}x^{2}(\lambda +\Delta^{2})} H_{n}(\Delta x) ,
\end{equation}
with the quantities $\lambda$ and $\Delta$ defined by
\begin{equation}\label{20}
\lambda=\frac{\beta -\alpha}{\omega-\alpha-\beta},\quad \Delta = \left (\frac{\Omega}{\omega -\alpha -\beta}  \right )^{\frac{1}{2}},
\end{equation}
and $H_{n}(\xi), \xi \in \mathbb{R}$ is the \(n\)th degree Hermite polynomial. The eigenfunctions are orthonormal with respect to the quantity $e^{\lambda x^{2}}$, i.e.,
\begin{equation}\label{21}
\int \psi_{n_1}^{*}(x)e^{\lambda x^{2}} \psi_{n_2}(x)dx=\delta_{n_1,n_2}.
\end{equation}
Following the same arguments as done in the context of the harmonic oscillator, an analogue representation for the quantized black hole area can be readily set up when we compare the representation (\ref{18}) with (\ref{9}). We find
\begin{align}
    \omega &= \alpha + \beta + \frac{8\pi}{m_P^2}, \label{22}\\
    \Omega &= \frac{8\pi}{m_P^2} ,\label{23}
\end{align}
in which $\alpha + \beta > 0$. Due to the presence of the parameters $\alpha$ and $\beta$, the above estimate of $\omega$ is different from what emerged in the case of the harmonic oscillator unless $\alpha$ and $\beta$ are of opposite signs. In fact, we see that in the non-Hermitian case, $\omega$ exchanges its role with $\Omega$ as is clear from (\ref{23}). A major motivation behind studying the area quantization of black holes within the framework of the Swanson Hamiltonian is that the Swanson Hamiltonian presents the most general quadratic-oscillator Hamiltonian and which also admits a spectrum that is isomorphous to that of the much-studied problem of the harmonic oscillator. Therefore, in a sense, one can obtain the results of the earlier developments (for example, of \cite{lou,jal,das2}) as a special case of the present analysis. Furthermore, as we shall see at the end of Sec. (\ref{BHsec}), our analysis provides a consistency check on the result of \cite{mitra} on the logarithmic corrections to the area law.

\section{Modified results for black holes}\label{BHsec}
The modified expressions above will profoundly impact the eigenvalue problem of the Hamiltonian. Corresponding to the kets $|n\rangle$, the eigenvalues will now reveal the spectrum $(n + \frac{1}{2})(\alpha + \beta + \frac{8\pi}{m_P^2})$, where $n = 0, 1, 2,\cdots$. Accordingly, the quantized relation for the black hole area would turn out to be 
\begin{equation}\label{24}
    A (\alpha, \beta) = \left (n + \frac{1}{2} \right ) \left (\alpha + \beta + \frac{8\pi}{m_P^2} \right ), \quad n= 0, 1, 2, \cdots.
\end{equation}
Note that the positivity of $\alpha + \beta$ is in tune with the area-increase theorem which requires $A(\alpha,\beta)$ to be a
non-decreasing function. Otherwise, the cosmic-censorship hypothesis could be violated leading to the possibility of the appearance of naked singularities in the universe \cite{penrose}. It may be pointed out that the estimate of $A(\alpha,\beta)$ is larger than the one given in (\ref{9}). On the other hand, the black hole mass is restricted to the discrete values

\begin{equation}\label{25}
    M (\alpha, \beta) = \mathcal{M} \sqrt{\Lambda}, \quad \Lambda = 1 + \frac{\alpha + \beta}{8 \pi} m_P^2 > 1.
\end{equation}
The expressions (\ref{22})-(\ref{25}) are novel results obtained in view of a non-Hermitian, \(\mathbb{PT}\)-symmetric representation of the area quantization. The main difference between (\ref{11}) and (\ref{25}) is that the latter contains an additional factor of $\sqrt{\Lambda}$ which implies that $M(\alpha, \beta)$ is numerically larger than the estimate of $\mathcal{M}$. \\

To derive the expressions for the temperature and entropy of the black hole in the present setting, we proceed in a similar manner as for the harmonic oscillator. The calculations are straightforward and we obtain the following expressions for $\mathcal{T} (\alpha, \beta)$ and $\mathcal{S} (\alpha, \beta)$:
\begin{align}
    \mathcal{T} (\alpha, \beta) &= \left(\frac{\kappa}{16\pi \sigma_S}\right)^{\frac{1}{4}} \frac{1}{\sqrt{\Lambda}} \frac{m_P^2}{2 \mathcal{M}}\left(1+\frac{1}{\Lambda}\frac{m_P^2}{16 \mathcal{M}^2 }\right), \label{26}\\
    \mathcal{S} (\alpha, \beta) &= \mbox{(constant)} A (\alpha, \beta) - \frac{1}{\sqrt{\Lambda}}\frac{\pi}{4}\ln A (\alpha, \beta) , \label{27}
\end{align}
the latter up to an integration constant. While analyzing the relative behavior between $\mathcal{T}(\alpha,\beta)$ and $\mathcal{T}$ we observe that the ratio is independent of the spectral quantum number `$n$' for large values of the latter according to $\frac{\mathcal{T}(\alpha,\beta)}{\mathcal{T}} \sim \frac{1}{\sqrt{\Lambda}}$, $n > > 1$. In other words, $\mathcal{T}(\alpha,\beta)$ is suppressed by a factor of $\frac{1}{\sqrt{\Lambda}}$ as compared to $\mathcal{T}$.  For $\mathcal{S} (\alpha, \beta)$, the interesting point to note is that suitable values of the parameters $\alpha$ and $\beta$ can be chosen so that logarithmic corrections can have appropriate coefficients like  $\frac{1}{2}$ or even other values (see, for example, \cite{mann, furs, page}). This gives a greater degree of flexibility in tracking down the entropy corrections. For instance, choosing 
\begin{equation}
\alpha + \beta = \frac{8 \pi}{m_P^2} \bigg( \frac{\pi^2}{4} - 1 \bigg)
\end{equation} gives logarithmic corrections of the form \(\mathcal{S}(\alpha, \beta) = \frac{A(\alpha, \beta)}{4G} - \frac{1}{2} \ln \big(\frac{A(\alpha, \beta)}{4G}\big) + \cdots\), consistent with \cite{mitra}; the coefficient $\frac{3}{2}$ of \cite{kaul}, however, may not be obtained from the above expression as this requires
\begin{equation}
\alpha + \beta = \frac{8 \pi}{m_P^2} \bigg( \frac{\pi^2}{36} - 1 \bigg),
\end{equation} which violates the condition \(\alpha + \beta > 0\). This supports the conclusion of \cite{mitra} which asserts that the coefficient \(\frac{1}{2}\) is more consistent than \(\frac{3}{2}\) based on the quantum-geometry formalism.\\

In fact, (\ref{27}) allows us to obtain a bound on the value of the coefficient of the area correction; if the corrections go as \(\mathcal{S}(\alpha, \beta) = \frac{A(\alpha, \beta)}{4G} - k \ln \big(\frac{A(\alpha, \beta)}{4G}\big) + \cdots\), then one should have 
\begin{equation}
\alpha + \beta = \frac{8 \pi}{m_P^2} \bigg( \frac{\pi^2}{16k^2} - 1 \bigg),
\end{equation}
and the positivity condition \(\alpha + \beta > 0\) now requires
\begin{equation}
|k| < \frac{\pi}{4}. 
\end{equation} This provides an upper bound on the magnitude of \(k\).

\section{Conclusions}
In this work, we analytically examined the area spectrum of a Schwarzschild black hole based on the standard approach of connecting its horizon area with the Hamiltonian of the harmonic oscillator, as a natural follow-up of the periodic nature of the canonical coordinate and its associated momentum, and also from the extended non-Hermitian perspective as modeled by the $\mathbb{PT}$-symmetric Swanson oscillator. In the former case, we analyzed Bekenstein's postulate of interpreting the black hole area as a quantized object that yields the eigenvalues of the harmonic-oscillator spectrum, while in the latter case, we presented a concrete construction of the explicit expressions of the mass, temperature, and entropy of the black hole in terms of the guiding non-Hermitian parameters. Finally, we examined the logarithmic corrections to the black hole entropy and found that it is possible to reproduce the earlier-known area-correction term \(-(1/2)\ln\)(area) by suitably choosing the parameters of the Swanson model.

\textbf{Acknowledgements:} B.B. thanks Brainware University for infrastructural support. A.G. is thankful to Akash Sinha, Chandrasekhar Bhamidipati, Sudipta Mukherji, and Narayan Banerjee for useful discussions and is supported by the Ministry of Education, Government of India in the form of a Prime Minister's Research Fellowship (ID: 1200454). S.S. is grateful to the Shiv Nadar Institution of Eminence for providing financial support and is also thankful to the Council of Scientific and Industrial Research (CSIR), Government of India [through Grant No. 09/1128(18274)/2024-EMR-I] for providing him with Direct-SRF fellowship. \\

\textbf{Data-availability statement:} All data supporting the findings of this study are included in the article.\\

\textbf{Conflict of interest:} The authors declare no conflict of interest.

\begin{widetext}

\appendix 

\section{Reduced action in Hamiltonian form}\label{app}
Let us summarize some key steps leading to (\ref{5}) (see \cite{jal} and references therein for more details). Consider a general spherically-symmetric ADM line element which goes as
\begin{equation}\label{1-1}
ds^2=-N(r,t)^2dt^2+\chi(r,t)^2(dr+N^r(r,t)dt)^2+R(r,t)^2d\Omega_2^2,
\end{equation}
where $d\Omega_2^2$ is the line element on the two-dimensional sphere of unit radius. Let us incorporate Kucha\v r's fall-off conditions \cite{kuc} which ensure that the coordinates $r$ and $t$ are extended to the Kruskal manifold, $-\infty< r,t<\infty$, and that the spacetime is asymptotically flat. 
Then the Einstein-Hilbert action functional in the Hamiltonian form shall read
\begin{equation}\label{1-2}
 \mathcal{I}=\displaystyle\int dt\int_{\Sigma_r}\Big\{ \Pi_\chi \dot\chi +\Pi_R\dot R-NH-N^rH_r\Big\}dr -\displaystyle\int\Big\{N_+M_++N_-M_-\Big\}dt,
\end{equation}
where \(\Pi_\chi\) and \(\Pi_R\) are the momentum densities conjugate to the fields \(\chi\) and \(R\), respectively; they read
\begin{equation}\label{1-3}
\begin{split}
\Pi_\chi=&-\frac{m^2_P}{N}R\left(\frac{dR}{dt}-\frac{dR}{dr}N^r\right),\\
\Pi_R=&-\frac{m_P^2}{N}\Bigg\{\chi\left(\frac{dR}{dt}-\frac{dR}{dr}N^r\right)+R\left(\frac{d\chi}{dt}-\frac{d}{dr}(\chi N^r)\right)\Bigg\}.
\end{split}
\end{equation}
Here, $\Sigma_r$ denotes the one-dimensional radial Cauchy surface which extends from $-\infty$ to $\infty$, and $H$ and $H_r$ are the super-Hamiltonian and the radial super-momentum constraints, respectively. As it was shown by Kucha\v r \cite{kuc}, solving the Hamiltonian and momentum constraints provides us an observable, namely, the mass ascribed to the spacetime, $0\leq M<\infty$; the corresponding conjugate momentum $-\infty<P_M<\infty$ reads as
\begin{equation}\label{1-4}
P_M=-\displaystyle\int_{\Sigma_r}\frac{\sqrt{\left(\frac{dR}{dr}\right)^2-\chi\left(1-\frac{2M}{m^2_PR}\right)}}{1-\frac{2M}{m^2_PR}}dr.
\end{equation}
This brings the action (\ref{1-2}) to the unconstrained Hamiltonian form
\begin{equation}\label{1-5}
\mathcal{I}=\int \Big\{P_M\dot M-(N_++N_-)M\Big\}dt,
\end{equation}
where the new time-dependent conjugate variables $(M,P_M)$ obey the Poisson bracket $\{M,P_M\}=1$. Following \cite{lou}, if we choose the right-hand-side asymptotic Minkowski time as the observer-time parameter, we should restrict to $N_+=1$ and $N_-=0$.  Then the reduced action (\ref{1-5}) coincides with (\ref{5}), where $H(M)=M$ is the reduced Hamiltonian.

\end{widetext}

\end{document}